\documentclass{sig-alternate-05-2015}
\usepackage[utf8]{inputenc}
\usepackage[english]{babel}
\usepackage{glossaries}
\usepackage{enumitem}
\usepackage{graphicx}
\usepackage{color}      % for comments
\usepackage{amssymb}    % for comments
\usepackage{csquotes} 
\usepackage{url}
\usepackage{fmtcount}
\usepackage{pbox}

\numberofauthors{6}
\author{
\alignauthor
Rinat Khatipov\\
      \affaddr{University student}\\
      \affaddr{Innopolis University, Russia}\\
      \email{r.khatipov@innopolis.ru}
\and
\alignauthor 
Manuel Mazzara\\
      \affaddr{Service Science and Engineering Lab}\\
      \affaddr{Innopolis University, Russia}\\
      \email{m.mazzara@innopolis.ru}
\and
\alignauthor
Aydar Negimatzhanov\\
      \affaddr{University student}\\
      \affaddr{Innopolis University, Russia}\\
      \email{a.negimatzhanov@innopolis.ru}
\and
\alignauthor
Victor Rivera\\
      \affaddr{Service Science and Engineering Lab}\\
      \affaddr{Innopolis University, Russia}\\
      \email{v.rivera@innopolis.ru}
\and
\alignauthor
Anvar Zakirov\\
      \affaddr{University student}\\
      \affaddr{Innopolis University, Russia}\\
      \email{an.zakirov@innopolis.ru}
\and
\alignauthor
Ilgiz Zamaleev\\
      \affaddr{University student}\\
      \affaddr{Innopolis University, Russia}\\
      \email{i.zamaleev@innopolis.ru}
}
\title{Hikester - the event management application}

\begin{document}

\maketitle

\begin{abstract}
Today social networks and services are one of the most important part of our everyday life. Most of the daily activities, such as communicating with friends, reading news or dating is usually done using social networks. However, there are activities for which social networks do not yet provide adequate support. This paper focuses on event management and introduces ”Hikester”.  The main objective of this service is to provide users with the possibility to create any event they desire and to invite other users. ”Hikester” supports the creation and management of events like attendance of football matches, quest rooms,  shared
train rides or visit of museums in foreign countries. Here we discuss the project architecture as well as the detailed implementation of the system components: the recommender system, the spam recognition service and the parameters optimizer.
\end{abstract}

\section{Introduction}
When the Internet has begun to be used by a wide audience social
networks started to appear and now they are growing at an alarm rate. For
instance, Facebook now has more than 2 billions active users every month.
There are different types of social networks which give people the ability to interact with each other in novel ways. Each of them has
a specific objective, for example communicating and sharing thoughts, or identifying alike people (with related algorithmic problems to solve such as computing trust and reputation of users \cite{MBGDMQN2013} or identify new possible connections relevant to our profile \cite{LebedevLRM17}). Regularly, new social networks appear to meet people needs. An area that we have found uncovered by the complex world of social application is the management of events and identification of suitable people for joint participation. We therefore attacked the problem and found solutions
developing an event management system: Hikester. The main idea of our
social network is to create a convenient interface for bringing people
together.
Social network requires a good architecture and design to build convenient
for users and well maintainable for developers web and mobile
applications. Also social network should be friendly for users. We achieve
that by developing an intuitive interface for interaction with application.
Our recommendation system is built to concern users needs and to be sure
that user won’t miss any event on which he most likely wants to go.
However, there are always people with evil intentions which must be
eradicated. Each social networks has the spam problem. An example is the
Nimses application, which began its work not so long time ago, but from
the first days they were not ready for spammers, and after several months of
work, people started leaving this application because of the large amount
of spam. In order to not repeat the mistakes of the Nimses, we decided to
develop a spam recognition system and then integrate it into the Hikester
application. All these components leads to a easy to use and convenience
social network.
\section{Related Work}
In the field of application architecture development has been done a lot of research, for example from \cite{design_patterns, Freeman-design-patterns} you can clearly deduce what design patterns are, how and for what purpose they are used. One of the most popular pattern in Web is MVC. MVC pattern, that implies the division of the project into 3 parts: Model, View, Controller. MVC design pattern can be applied for Hikester project. Another modern way to develop project is following the principles of liquid architecture. This structure which satisfies the requirements of the manifesto described in the article \cite{liquid-software}. 

If we add the conditions of the liquid architecture to the MVC, then we will add the condition for Model: mobile and web Models will be used a single storage. So all user's interacts with the system will done through a single repository. Such storage could be real-time database - Firebase \cite{firebase-doc}. Which ensures that synchronization of the full state of application and the data is displayed correctly when using different devices. This allows to users "be able to effortlessly roam between all the computing devices that they have"\cite{liquid-software}. More information about how to built liquid software can be found in \cite{liquid-js}. To support liquid architecture, another important requirement for the front-end is responsive design. About this is written in the article \cite{towards-liquid-soft}.

Asynchronous method invocation also is design pattern which described in \cite{async-lavender}. The servers which use the asynchronous requests could be scalable in more easily way, in contrast to the typical implementation based on the threaded model. For example, in \cite{node-about} described Node.js server. Also node.js and its novel implementation \cite{node-DPWS} are very good at coping with small data (e.g. sensing data).

Using new software approach in web's front-end - Single page application, described in \cite{spa,spa-2}. Based on this paradigm were created several front-end frameworks, one of them is a reactJS \cite{react-mastering, react-doc}. 
\section{An Event Management System:\\ Hikester}
\subsection{Project requirements}
To make the event management application more convenient for users, we will apply the following approaches:
\begin{itemize}
\item \emph{Real time application -} that approach assume all data updates will be shown in real time. For example if some person create event, that event will be immediately shown on the screen of all other users that looking in the same location where event was create. 
\item \emph{Single page application (SPA) -} application which built by that approach should avoid interruption of the user experience between successive pages. SPA represent the application more like a desktop application. User`s browser just once load all necessary code. And all users`s interactions with single page application will dynamically change just some part of application without any page reloading at any point in the process, nor does control transfer to another page. 
 
\item \emph{Cross-platform application -} the application should work correctly on various types of devices such as a PC, tablet and phone. The number of users will directly proportional to the coverage of devices for which this application will be implemented. 
\end{itemize}

\subsection{Design/Architecture}
\subsubsection{Main components}
For the application to be effective, it is necessary to develop a user-friendly interface. That application should not restrict users in choosing the device through which they would like to use the Hikester application. With this in mind, the software design components of Hikester application consist of the following parts:

\begin{itemize}
\item \emph{Web interface} 
\item \emph{Mobile applications (Android \& IOS)} 
\item \emph{Real-time Firebase database\cite{firebase-doc}} 
\item \emph{Back-end} 
\end{itemize}

That main components we can divide into 3 groups: 
\begin{itemize}
\item \emph{The Web and mobile applications - responsible for representation of data for user. Also that components will send to back-end some user interactions. } 
\item \emph{The back-end - responsible for data control and process the user interaction.} 
\item \emph{Real-time Firebase database - represent data.} 
\end{itemize}

That was the reason to choose MVC (Model View Controller) pattern as the main design pattern for project in general. That is one of the most popular pattern in modern web development.\cite{mvc_the_best} 

\begin{figure}[h]
	\centering
    \includegraphics[width=0.5\textwidth]{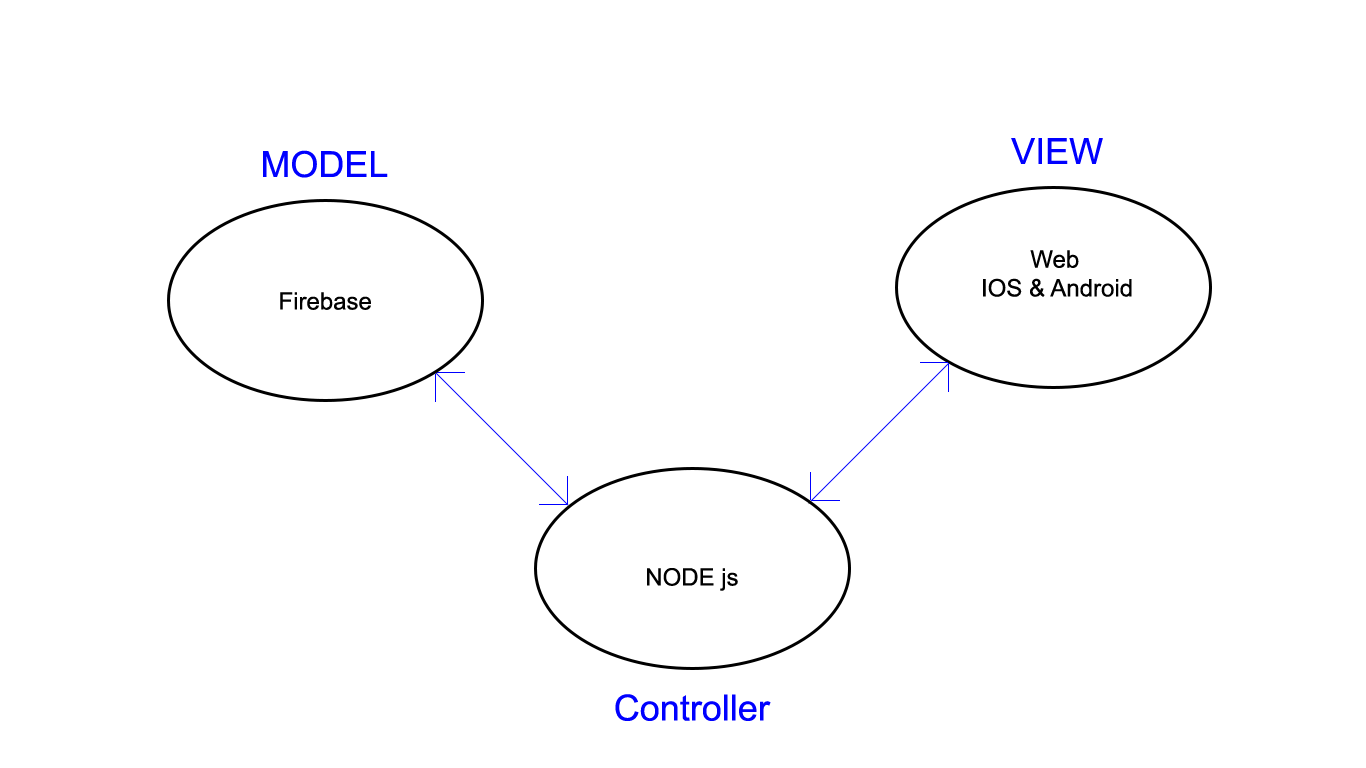}
	\caption{Project components interaction.}
	\label{fig:interaction}
\end{figure}

In context of that project Web and mobile applications show to users all information which was received from model (real-rime Firebase database). That two project components corresponds to View part of MVC. The back-end handle all users interactions that was sent from View and change the data in accordance with the action which was done by user. It represent Controller in terms of MVC pattern. 
The real-time Firebase database directly manages the data. Also it is responsible for logic and rules of the writing and deleting the data independent of the user interface. That component is the Model in MVC pattern.

Figure~\ref{fig:interaction} illustrate project components interaction. 

This structure partially satisfies the requirements of the liquid architecture manifesto. The liquid architecture relate to applications that have several implementations for example it has web and mobile versions. Also that approach assume that the application should have a data synchronization in the context of one application especially when one user will use several devices and as a result several implementations of that application. And each of that implementations should represent user's data in a correct way. It means the date should be the same, but it can be shown in a different ways. All of this requirements are necessary to make the application more convenient for the end-user. 

In the Hikester application, all user information and data stored in single repository, namely the Firebase real-time database, which ensures that synchronization of the full state of application and the data is displayed correctly if the user will need to use several devices to use the application. This allows to users "be able to effortlessly roam between all the computing devices that they have"\cite{liquid-software}.

Another approach that was used to support liquid architecture - responsive web design (RWD). RWD need to render well all web page data on a different devices regardless of the screen size and type of device. The main idea for the responsive web design was in creating several views for one web page. These web page views depended on the current screen size of the device from which the page display request was came. And the modern stylistics of the web page - Cascading Style Sheets (css) allows you to achieve RWD using a minimum of effort for this.
\subsubsection{Auxiliary components}
\begin{figure}[h]
	\centering
    \includegraphics[width=.5\textwidth]{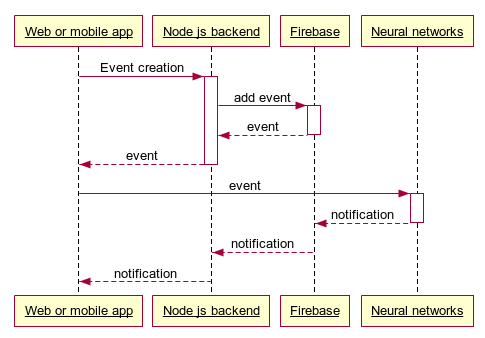}
	\caption{Project modulus interaction sequence diagram.}
	\label{fig:data_flow}
\end{figure}

For more comfortable work with main components and for achieving additional functionalities such as recommendations for users to join for some events and spam recognition system Hikester project also has separate auxiliary services:
\begin{itemize}
\item \emph{Elasticsearch} 
\item \emph{Geofire} 
\item \emph{Python neural network} 
\end{itemize}

In general, all modulus interactions take place according to following schema: 

All user interaction are handled by back-end. After processing the user's input back-end send request to Firebase database. Firebase manage data and send response for back-end. The back-end in its turn send it to View (Web or mobile application). After that manipulation the neural networks is notified about some changes. And after processing data neural networks interact with real-time Firebase database. Then it send it to back-end and then user get some notifications, which was sent by neural network. The figure~\ref{fig:data_flow}  shows example of that interactions.
\subsection{Implementation}
\subsubsection{Back-end}\label{FR}
The back-end of project implemented by using “Node.js”. In order to provide a convenient way to interact with front-end and mobile applications the back-end was implemented as an API with Express framework. The main job of back-end is to communicate with front-end and mobile applications via JSON requests.

Node.js is one of the newest platforms based on the V8 engine. That engine translate JavaScript into machine code that promotes to turns JavaScript from a highly specialized language into a general-purpose language. Due to the fact that Node.js use the asynchronous requests and the most part of that platform is implemented in languages C and C++ Node.js provides quite a high speed for input/output (i/o) operations. Also each new connection requires a small memory area in the heap. All this benefits allow to scale the application which based on the Node.js in more easily way, in contrast to the typical implementation based on the threaded model.

However, the node is not doing so well with processing large static files \cite{is-node}, but this problem is becoming less noticeable with the release of new versions. Although in that project case, Hikester application works only with small data, such as geolocation, message, recommendations, but in large amount. For that kind of data the node.js and its novel implementation such as node-DPWS \cite{node-DPWS} will be one of the best solutions from the range of modern existing software development platforms. In addition the node js server can be integrated in firebase platform as cloud functions that will be covered in firebase subsection. That integration allows to scale application in the easiest way. In that case all computing power will depend only on the payments that will be paid to the firebase platform.
\subsubsection{Firebase}\label{Methods}
Firebase is a platform for development mobile and web application. Real-time nosql database is one of the main service of that platform through which all components of the event management application will interact. The service provides an API that allows application data to be synchronized across clients and stored in Firebase database. There is enable integration with Android, iOS, JavaScript, Java, Objective-C, Swift, React and Node.js applications. All of that components are used in Hikester application. The real-time database can be secured by using the special server-side-enforce security rules that provided by firebase.

\begin{figure}[h]
	\centering
    \includegraphics[width=0.5\textwidth]{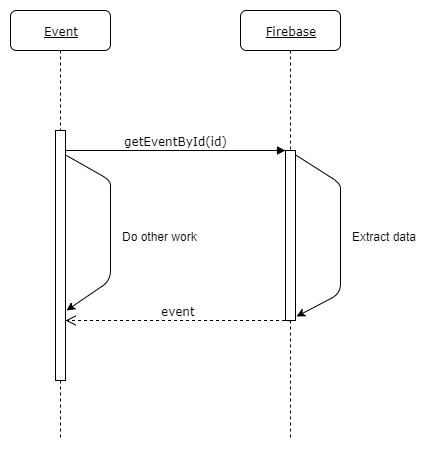}
	\caption{Firebase sequence diagram.}
	\label{fig:firebase}
\end{figure}

All requests to firebase real-time database for add or extract data are executed asynchronously. Example of such request illustrated in figure~\ref{fig:firebase}. There is shown request from back-end JavaScript Event object to Firebase database to extract certain event by its id. While Firebase will prepare the data for send it back, the back-end able to execute some other part of code. And when the data will be accepted by the Node.js server, it can continue execute the part of code which was interrupted by request to firebase database.

Another useful tool that provided by Firebase is cloud functions. Cloud functions is JavaScript code that execute on special cloud firebase environment, that can handle a HTTP or HTTPS requests. That functions allow to use simple npm modulus and can be written as simple Node.js server. Also they can constitute a trigger for real-time database that executes when the data is changed. 

Cloud functions provide a connection between neural networks and the user side components in the context of event management application. For example, when user create an event, the trigger for creating the record executes and the necessary event's data is entered into another special table, which is listened to by neural networks. Having made the necessary data processing, neural networks decide that to do some action. For example as result neural networks will again create a record in the firebase database, for example it can be some notification for the user or detection that there is spam event.

Also, firebase allows you to collect analytics of the entire application. That analytics provides insight into application usage and user engagement. And such kind of analytics data will help to find the vulnerabilities in the application, find which part of application is more convenient to users, which needs to be finalized. Also that data will contribute to the creating a marketing campaign in the future, which in turn will stimulate the activity of old users and  should be able to attract new users.

\subsubsection{Front-end web }\label{example}
Web part of Hikester implemented as a single page web application. The logic of the front-end implemented on the JavaScript, using the reactJS framework, and the web application state control system - REDUX. The principle of single page application is achieved thanks to the state model of the ReactJS framework. That model allow to handle changing of some attribute of virtual DOM element in certain moment of time and when it was changed the ReactJS change a state of certain element on web page. So the web page can change each element that contains without reloading. That is the reason why the event management application is single page application. 

Front-end communicate with back-end and firebase database through asynchronous request. For more convenient way to work with such kind of request used Promises principle of JavaScript. The connection with the back-end implemented via JSON API using a third-party library - Axios. To support liquid architecture, another important criterion for constructing the front-end was the availability of responsive design.

\subsubsection{Mobile application (IOS \& Android) }\label{example}
The key task of developing a mobile application is to make the most simple, convenient and user-friendly interface for the user. 

The main design pattern on which built mobile applications based on is VIPER: View, Interactor, Presenter, Entity, and Routing. That pattern allow to develop applications on high level of abstraction. Separating application on several different and independent part will provide to more easily support in future. The VIPER design pattern allows to achieve such kind of separation. 

\begin{figure}[h]
	\centering
    \includegraphics[width=0.5\textwidth]{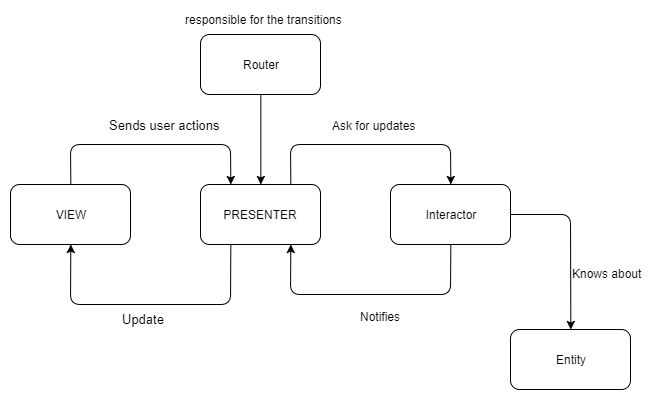}
	\caption{VIPER. Design pattern's modules interaction description.}
	\label{fig:viper}
\end{figure}

Figure~\ref{fig:viper} illustrate the interaction of that components. All interactions with mobile applications follow the next schema:

When user do some action on the View component, it send request to Presenter. Then Presenter recognize what the action was, some input/output operation, confirmation, event creation etc. And then it decides which step will be next. Presenter component asks for Interactor to updates some data or not. If it is necessary Interactor will change the data of Entity component and after all changes will send it back to Presenter. The Entity manage the data in firebase real-time database. When control comeback to Presenter that will change some data or state of View. This is the sequence of the steps that mobile application's do for each response for some action that came from the user. The Router components are responsible for all the interactions between all other components. 

As the front-end mobile applications connection with the back-end was maintained using the JSON API.

The mobile applications use Observable design pattern for communication with Firebase real-time database. 

\begin{figure}[h]
	\centering
    \includegraphics[width=0.45\textwidth]{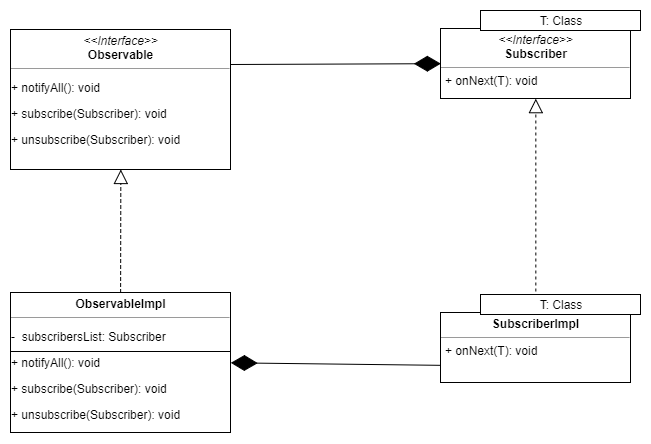}
	\caption{Observable design pattern UML diagram.}
	\label{fig:observer}
\end{figure}

That design pattern's UML diagram is represented in figure~\ref{fig:observer}. Observable design pattern consists of 2 interfaces - Observable and Subscriber. Also it has theirs 2 implementations. The main idea is that all observable objects contains some list of subscribers and when there is happen some action with Observable object it will notify all subscribers about it. Also Observable able to subscribe some new subscriber or unsubscribe some old one. Subscribers have onNext(T t) method that allow to get some object as result of update, and then do some computations with that accepted date. So when action is happen with Observable object it call onNext(T t) method on all subscribers and send to them some new data.

Firebase real-time database represent the observable object and all users request is related to firebase as subscriber. For example if user A is searching some event and filtered the searching by certain tag and time, the application will handle the firebase database for some changes and will extract all necessary data. And if at that moment some user B will create a event with parameters that satisfy the user A searching criteria the firebase real-time database (observable) will notify all users (subscribers) including user A that there was some change. And user A will see in real time that there will appear some new event in which he may be interested without any additional user actions.

\subsubsection{Auxiliary services}\label{example}

\subsubsection*{5.1}{Neural networks}\label{UC}

The neural networks in Hikester project is represented as a separate service, which is accessed through back-end and firebase cloud functions. The need for neural networks is due to the fact that, for the convenience of the user, it is necessary to create a more user-friendly environment. That will ultimately determine the user's needs by itself, as well as shield users from unwanted content. 

There are 3 neural networks:

\begin{itemize}
\item \emph{Recommendation system} 
\item \emph{Spam recognition system} 
\item \emph{Python neural network} 
\end{itemize}

For implementing all that features were developed neural networks. These networks are designed on Python, with using special neural networks  libraries.

\paragraph{Recommendation system}\label{UC}

This systems is responsible for analyze events which was recently created. The main data process of recommendation system identify the interests of users.

In Hikester project each user is related to some group. That group identified by user social profile and also by behavior in event management application, such as creating certain type of event, accepting  or refusal of participation on some events. When user create some event, the neural networks is notified about it and start analyze that data. After that recommendation system send notifications to all users which may be interested on event which was recently created.

\paragraph{Spam recognition system}\label{UC}
This systems is responsible for analyze the content of event which was recently created. Spam recognition is reasonably important because prevent users from malicious people who want to spoil the application and from seeing a spam in the application is one of the most relevant part in each project. It is very uncomfortable to use any application when there are a lot of low-quality content.

\paragraph{Parameter optimizing system}\label{UC}

The Parameter Optimizing System suggest to users, who create some new event, which parameters (date, time, location, tags) is more relevant to choose, for achieving the bigger activity and greater participation of people on that event. The system able to do some decision which parameters should be chosen in specific case. That decision based on  processing the data of previous events.

\subsubsection*{5.2}{Elasticsearch}\label{UC}
The Elasticsearch search engine is used to implement a full-scale search in noqsl data base. This tool provides a distributed, multitenant-capable full-text search engine with an HTTP web interface and schema-free JSON documents.

\subsubsection*{5.3}{Geofire}\label{UC}
Geofire allows to use real-time location queries with firebase real-time database. This tool needs for the display of certain events in a given radius from the current user's or given location.
\subsection{Features}
\subsubsection{Spam Filtering}
Social networks are growing at an alarming rate, and networks such as Facebook
and Twitter has more then 100 million users a month and due to this
they have become main means of communication. This attracts spammers
to social networks and leads to increasing the amount of social spam. Therefore, every
social network must create its own independent spam filter.
The existing methods of detection can be divided in the context of machine learning
into supervised and unsupervised. Secondly, they can be divided into three categories according to their characteristics: behavioral, linguistic or those
that use a combination of these two. 
 in order to overcome
the problem of spam, a lot of research has been done and various methods of
spam filtering have been implemented. A spam filter is a set of instructions
for determining the status of a received feedback, a letter on the mail. The
main task is how to create an effective spam filter that allows you to skip
the necessary information by blocking unwanted information. Unfortunately,
there is no approach to spam filtering, which guarantees a 100 procents spam
free environment to this day. This is because every anti-spam filtering approach
has its own limitations.
There are three main measures of spam filters quality: efficiency, accuracy
and simplicity of administration. Efficiency refers to the percentage
of recognized spam messages. The task of spam recognition is complicated
for many reasons. Even a person can sometimes not easily distinguish spam
from necessary information. There is also the problem of aggressive pursuit
of efficiency, which leads to the blocking of normal information. Spammers
improve their skills, and today’s defense can quickly become obsolete, spammers
will find other ways to bypass protection. Still another important thing,
as false triggers increase, recipients lose confidence in blocking spam. When
users think that they can not rely on filtering accuracy, they are forced to
manually separate spam from the necessary information.
Now a huge number of different algorithms are used to recognize spam.
One of the most successful algorithms is the Naive Bayes classifier. Naive
2
Bayes classifiers are a popular statistical technique of e-mail filtering. They
typically use bag of words features to identify spam e-mail, this approach is
commonly used in the text classification.
Naive Bayes classifiers work by correlating the use of tokens, with spam
and non-spam e-mails and then using Bayes’ theorem to calculate a probability
that an email is or is not spam.
Naive Bayes spam filtering is a baseline technique for dealing with spam
that can tailor itself to the email needs of individual users and give low false
positive spam detection rates that are generally acceptable to users. The
Bayes algorithm is the most popular algorithm in spam recognition, but there
are also omissions in it. For example, you need to have a dataset of good
and bad messages to initialize the filter. Which is unacceptable at the startup
stage of the Hikester project, since in order to type a training sample,
you need to have a huge number of verified users. Also another important
problem of this algorithm is poor performance. On each message, a userspecific
database of the word probabilities has to be consulted. What we
can not use this approach in our project, because the Hikester application is
completely real-time.
Almost all algorithms for spam recognition are reduced to classification
problems, one of them is K Nearest Neighbors Classical. The main principle
of this algorithm is to attach messages to certain groups, based on what
messages are ”around it”. we may try to classify a message according to the
classes of its nearest neighbors in the training set.
Also, artificial neural networks are used to detecting spam Artificial
neural networks (ANN-s) is a large class of algorithms applicable to classifi-
cation, regression and density estimation [10].One of these neural networks
is Perceptron. The idea of a perceptron is to find a linear function of the
vector function $f (x) = w
T x + b$ such that $f(x) > 0$ for vectors of the same
class, and f(x) < 0 for vectors of another class.
In addition to the perceptron, Multi-Layer Perceptrons(MLP) are much
more often used in spam recognition systems[10]. In MLP, each hidden neuron
is equivalent to a perceptron. The main feature of MLP for the Hikester
application is that this neural network is able to generalize to new date. It
can help solve the problem of spam in the startup phase.
Also, one of the most popular methods for recognizing spam in mail is
the Support vector machine (SVM). SVM is a relatively new method for
classifying both linear and nonlinear data. SVMs are supervised learning
models that are associated with learning algorithms that analyze data and
recognize patterns. The primary SVM takes a set of input data for each
given input, which has two possible form classes, which makes it an unvariant
binary linear classifier.

\subsubsection{Recommender System}

\begin{figure}[]
	\centering
    \includegraphics[width=0.5\textwidth]{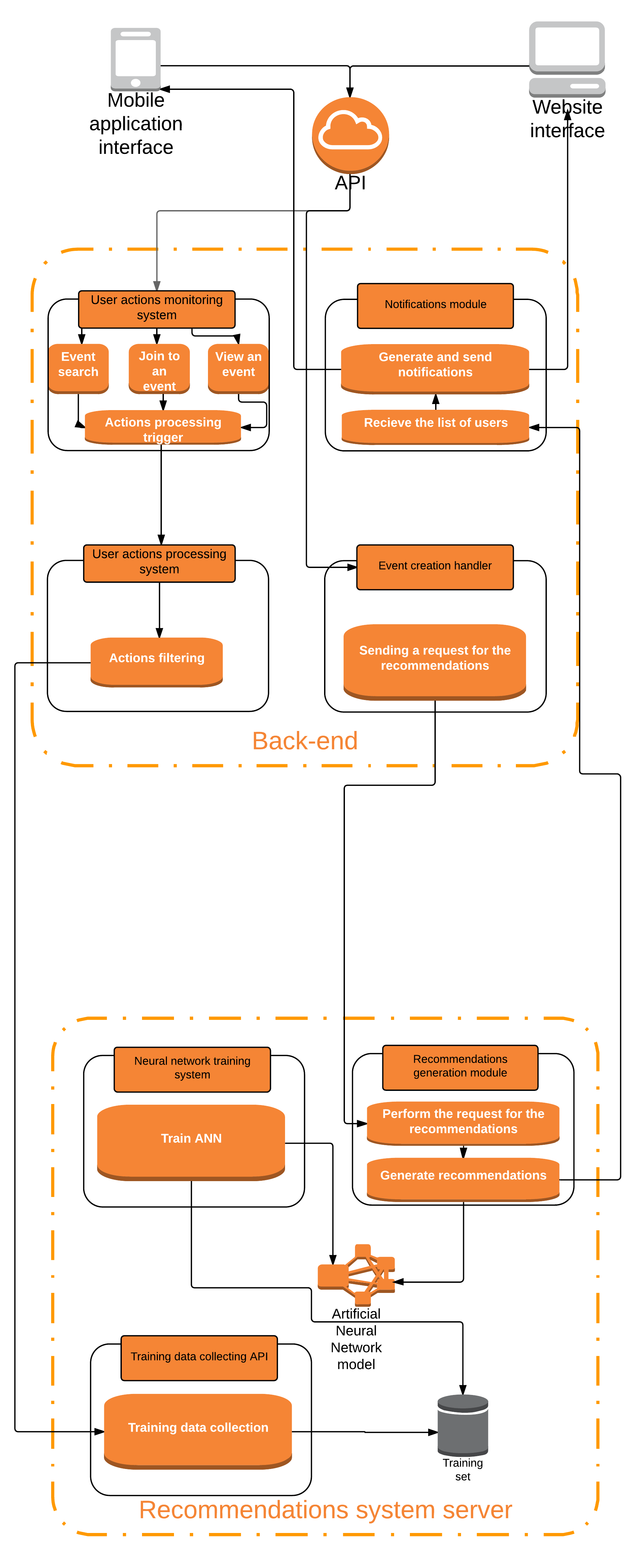}
	\caption{Recommendations system architecture}
	\label{fig:interaction}
\end{figure}
The growth of new social networks is very fast. Due to this, a huge amount of content is generated in them. But users of a social network have different needs. For an ordinary user of a social network is very difficult to find useful information for him in such a variety of content. And speaking about the Hikester project, in which users are supposed to choose activities which they are interested in, the variety of content is particularly difficult. The solution discussed in this chapter involves the creation of an intelligent system for generating proposals based on neural networks and machine learning algorithms. The main idea of this system is to isolate the special features of each user through trained neural networks to further formulate an offer for an event or activity that best meets the user's needs.
In order to provide for users an easy way to find the activities they are interested in we need the offers system. The purpose of this system is to attract new participants to the event by notifying them of new events that might be interesting to them.
The user interacts with the recommender system through the interface of the website and mobile applications. The interface allows user to create an event, register on an existing event, search for a specific event using various filters, view information about an existing event, communicate with other event users, cancel their participation in the event. User actions such as finding an event using filters, viewing an event, and joining an event trigger the  recommendation system module. Each such triggering is responsible for filling the sample for the subsequent retraining of the neural network for the recommendation system. Such an action as the creation of a new event calls the function of the second module of the recommendation system: the Recommendation Generation module. This module sends the information about a new event as an input to a neural network and then receives data in the form of a list of users from the neural network for which this event may be interested in and for each user from the received set forms a request to send a notification of the event to the user. In order to give users high-quality events recommendation system, the system should work accurate. In order to achieve maximum accuracy, the input data must be selected in the right way, and the amount of new input data that is required to retrain the neural network should be selected as efficiently as possible.

\subsubsection{Parameter Optimizing System}
The amount of possible events created in event management system is huge. And there’s need of each person who wants to create an event to make his occasion more popular and get larger amount of people if he wants to. People may try to think a little bit and find out in what days, at what time and where this particular event will be more appropriate to be created. But machines do this task better. 

The parameter optimizing system helps users, who want to create an event, in choosing the most optimal event parameters to make it more popular to other people who use the event management application. This system analyses previous application event history and decides what event parameters would be more optimal to choose in specific case.

After the analysis of which of the all possible event parameters could be optimized and then suggested to users by the parameter optimizing system it was decided that there are three main event parameters which could be analyzed better by the computer comparing to the human because of the computation power: time, date and location.

The system has the following components:
\begin{itemize}
\item Two neural networks for retrieving the most optimal values for two different event parameters:
\begin{itemize}
\item Time
\item Date
\end{itemize}
\item The algorithm for detecting the most popular places for the event with specific tag, time and date
\item User activity handler
\item Neural network training component
\end{itemize}

\begin{figure}[h]
 \centering
    \includegraphics[width=0.5\textwidth]{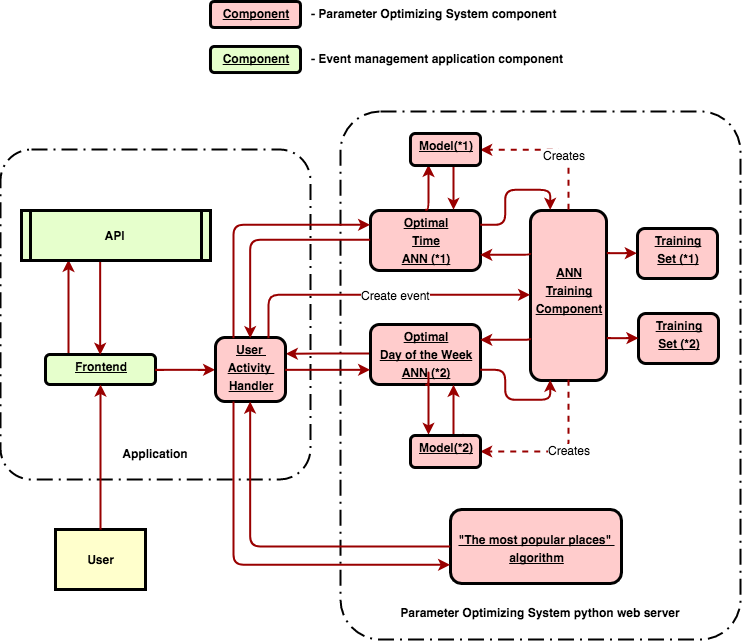}
 \caption{Parameter optimizing system components interaction}
 \label{fig:fig_pos_components_interaction}
\end{figure}

\subsubsection{User activity handler}
The user activity handled is used for monitoring user activity in the applica- tion and to establish a connection between the user actions in the front-end and the parameter optimizing system. There are four types of situations where user activity handler will be executed:

\begin{itemize}
\item User created an event and then information about the event sends to ANN training component to retrain the model
\item User needs help in choosing the most optimal time. User activity han- dler sends request to the “Optimal Time” ANN to get prediction
\item User needs help in choosing the most optimal date. User activity han- dler sends request to the “Optimal Date” ANN to get prediction
\item User wants to know where are the best areas to create an event. User activity handler sends request to the “The most popular places” al- gorithm to get several places where holding the event will be more efficient
\end{itemize}

\subsubsection{Neural network training component}
Neural network training component is the core part in the parameter op- timizing system. It generates neural network models for “Optimal time” ANN and “Optimal date” ANN from the corresponding training sets. When the component receives request from the user activity handler it inserts new training tuple into the training sets. The training component generates new neural network model when the training set has specific amount of new tu- ples. This will make models actual at every moment neural networks use them.

% \subsubsection{“Optimal time” ANN}

\section{Validation}
The requirements which was mentioned in chapter 3.1 was done in the current state of application. This was facilitated by the fact that the main platform which was chosen for developing the application was the Firebase platform. The greatest advantage of using this platform is the availability of a Firebase real-time database. That allowed to built real-time application in the most easiest way. This kind of user interaction with the system makes it possible to provide a very convenient user interface. Thus ensuring a very fast and comfortable using of Hikester application.

Also event management application satisfy to requirement for single page application. The ReactJS framework provided state approach for that project. Each user interaction change just part of web page. User only need to download once all the data - HTML, JavaScript and css. After that the application become like desktop application. Another feature of that approach is if there will be some problems with domain or hosting but other components of application will work and user will have downloaded page in browser, the application will work like there is no any problem.

For achieving the cross-platform approach there were developed two mobile applications for iOS and Android operation systems. That two applications are very similar. The main differences between them only in styles. That approach provide the coverage to a wider audience. So user can use any  modern device, leading in the field of mobile devices. Also responsive web design provide that opportunity in more easily way, if user want to use just web site he can do it without any restrictions on device.
\section{Conclusions and Future Work}
In this section we will discuss the application's additional functionality to be implemented in future as well as the main conclusions of this work. One of the main parts of the application which is not implemented yet is the mobile version. The mobile app will be the main tool to use "Hikester". We believe that the mobile version of our application will become the most convenient and effective way to create, search and join an event. 

From the recommender system perspective, the one of the most significant features which we need to implement is the Location-Based Recommendations System. The main idea of this feature is to divide people to the locational groups and generate for them group offers which will be based on the most popular event types within the current location. This will give us the possibility to take into consideration one of the most important information: the location of user, in order to generate a recommendations for him.
    
There are several other features to be implemented in future, but now the mobile applications and the Location-Based Recommender System are the top priority tasks. It is particularly interesting to understand how the intelligent service provided by Hikester can be integrated with other intelligent system that we are investigating, for example in the field of smart buildings \cite{Salikhov2016a,Salikhov2016b} or in the automotive sector \cite{Gmehlich13}, either for recreational purposes or for assistance to people in need \cite{Nalin2016}.

The main task in creating the "Hikester" application was to provide users with a convenient and understandable platform for creating, searching and joining an event. During the development of this application, a large amount of additional functionality like a recommender system,  parameters optimizing and spam detection modules has appeared. Due to such features the project became more convenient and attractive for users. However, at this stage there are still many problems that we need solve. We believe that in the nearest future "Hikester" will become a part of our everyday life and will bring a great value to the society.

\bibliographystyle{unsrt}
\bibliography{bibl}

\end{document}